\documentclass[aps,prb,twocolumn,floatfix]{revtex4-2}
\usepackage[utf8]{inputenc}
\usepackage[english]{babel}
\usepackage{mathtools}
\usepackage{newtxtext}  
\usepackage{amsmath}
\usepackage{amsfonts}
\usepackage[varbb,varg]{newtxmath} 
\usepackage{bm,dsfont,eucal}
\AtBeginDocument{\renewcommand{\vec}[1]{\bm{#1}}}

\usepackage[dvipsnames,svgnames,x11names,hyperref]{xcolor}
\usepackage{hyperref}
\hypersetup{ 
    colorlinks=true, linkcolor=NavyBlue, urlcolor=NavyBlue, citecolor=NavyBlue
}
\usepackage{physics,braket,siunitx,slashed}
\usepackage[capitalise]{cleveref}

\begin{document}
        \title{Quarter-Metal Phases in Multilayer Graphene: Ising-XY and Annular Lifshitz Transitions
} 

\author{Mainak Das and Chunli Huang}
\affiliation{Department of Physics and Astronomy, University of Kentucky, Lexington, Kentucky 40506-0055, USA}

\date{\today} 

\begin{abstract}
Recent experiments have uncovered a distinctive magnetic metal in lightly-doped multilayer graphene, coined the \textit{quarter metal}. This quarter metal consolidates all the doped carriers, originally distributed evenly across the four (or twelve) Fermi surfaces of the paramagnetic state, into one expansive Fermi surface by breaking time-reversal and/or inversion symmetry. In this work, we map out a comprehensive mean-field phase diagram of the quarter-metal in rhombohedral trilayer graphene within the four dimensional parameter space spanned by the density $n_e$, interlayer electric potential $U$, external magnetic field parallel to the two-dimensional material plane $B_{\parallel}$ and Kane-Mele spin-orbit coupling $\lambda$. We found an annular Lifshitz phase transition and a Ising-XY phase transition and locate these phase boundaries on the experimental phase diagram.
The movement of the Ising-XY phase boundary offers insights into $\lambda$. Our analysis reveals that it moves along the line $\partial n_e/\partial B_{\parallel} \sim -0.5\times 10^{11} \text{cm}^{-2}\text{T}^{-1}$ within the $n_e$-$B_{\parallel}$ parameter space when $\lambda=30\mu$eV. Additionally, we estimated the in-plane spin susceptibility of the valley-Ising quarter-metal  $\chi_{_\parallel}\sim 8~\mu\text{eV} ~\text{T}^{-2}$. Beyond these quantitative findings, two general principles emerge from our study: 1) The valley-XY quarter metal's dominance in the $n_e-U$ parameter space grows with an increasing number of layers due to the reduce valley polarization variations within the Fermi sea. 2) Layer polarization near the band edge plays an important role in aiding the re-entrance of the paramagnetic state at low density. The insights derived from the quarter metal physics may shed light on the complex behaviors observed in other regions of the phase diagram.


\end{abstract}

\maketitle
\section{Introduction } 
The recent discoveries of symmetry-broken electronic phases \cite{zhou_half_2021,zhou2022isospin,zhou_superconductivity_2021,seiler2022quantum,han2023orbital} in the density-interlayer potential ($n_e-U$) parameter space of multilayer graphene have opened new avenues to explore strongly correlated physics \cite{han2023correlated,du2021engineering,de2022cascade,zhang2023enhanced,doi:10.1021/acs.nanolett.3c01262,10.1088/1674-1056/acddcf,lee2014competition,pantaleon2022superconductivity,holleis2023ising,lu2023fractional} and energy-efficient electronic devices \cite{PhysRevLett.95.226801,RevModPhys.92.021003,han2014graphene,zhumagulov2023emergent,huang2023spin,jimeno2023superconductivity,Katti_2023,PhysRevB.85.115423,gmitra2009band,ochoa2012spin,boettger2007first,imura2010anti,PhysRevLett.122.046403,PhysRevLett.124.177701,morissette2022electron}. The observed phases include 
spin-singlet superconductor, spin-polarized superconductors and generalized metallic ferromagnets, where the order parameter is characterized by the 15 generators of the $SU(4)$ space spanned by spin ($s$) and valley ($\tau$) degrees of freedom. 
Among the various magnetic phases, the so-called quarter metal is particular noteworthy because of its simplicity and its analogies to the relatively well-understood $\nu=\pm1$ graphene quantum Hall magnet. 
In the large hole-density regime of the quarter metal, the area enclosed by the Fermi surface extracted from magnetic oscillation data precisely corresponds to $4\pi^2$ times the carrier density. This implies that, despite the 4-fold degenerate density-of-states implied by time-reversal and inversion symmetry, all holes are enclosed by a single Fermi surface. Progressing to lower hole density in this phase, Zhou.~et.~al \cite{zhou_half_2021} found a Lifshitz transition where the simple Fermi sea becomes an annular Fermi sea. Despite its extensive capability, magnetic oscillations on its own does not provide insight into the nature of the order-parameter. Indeed, the order parameter can rotate in 15 the dimensional space without altering the area enclosed by the Fermi surface(s). Such variations in the order-parameter leads to variations in groundstate energy, which we found to be much smaller than the Coulomb energy scale. We termed this energy landscape as magnetic anisotropy energy landscape, and it holds pivotal importance in the study of magnetism. The intrinsic spin-orbit coupling (SOC) of graphene profoundly influences this magnetic anisotropy energy landscape. Since Kane and Mele first introduced it \cite{PhysRevLett.95.226801}, numerous experiments have employed varied techniques to measure the SOC \cite{arp2023intervalley,banszerus2021spin,island2019spin,wang2023electrical,koh2023correlated,PhysRevB.107.L201119,PhysRevB.81.241409}. 


\begin{figure}[t]
    \centering
    \includegraphics[width=0.9\columnwidth]{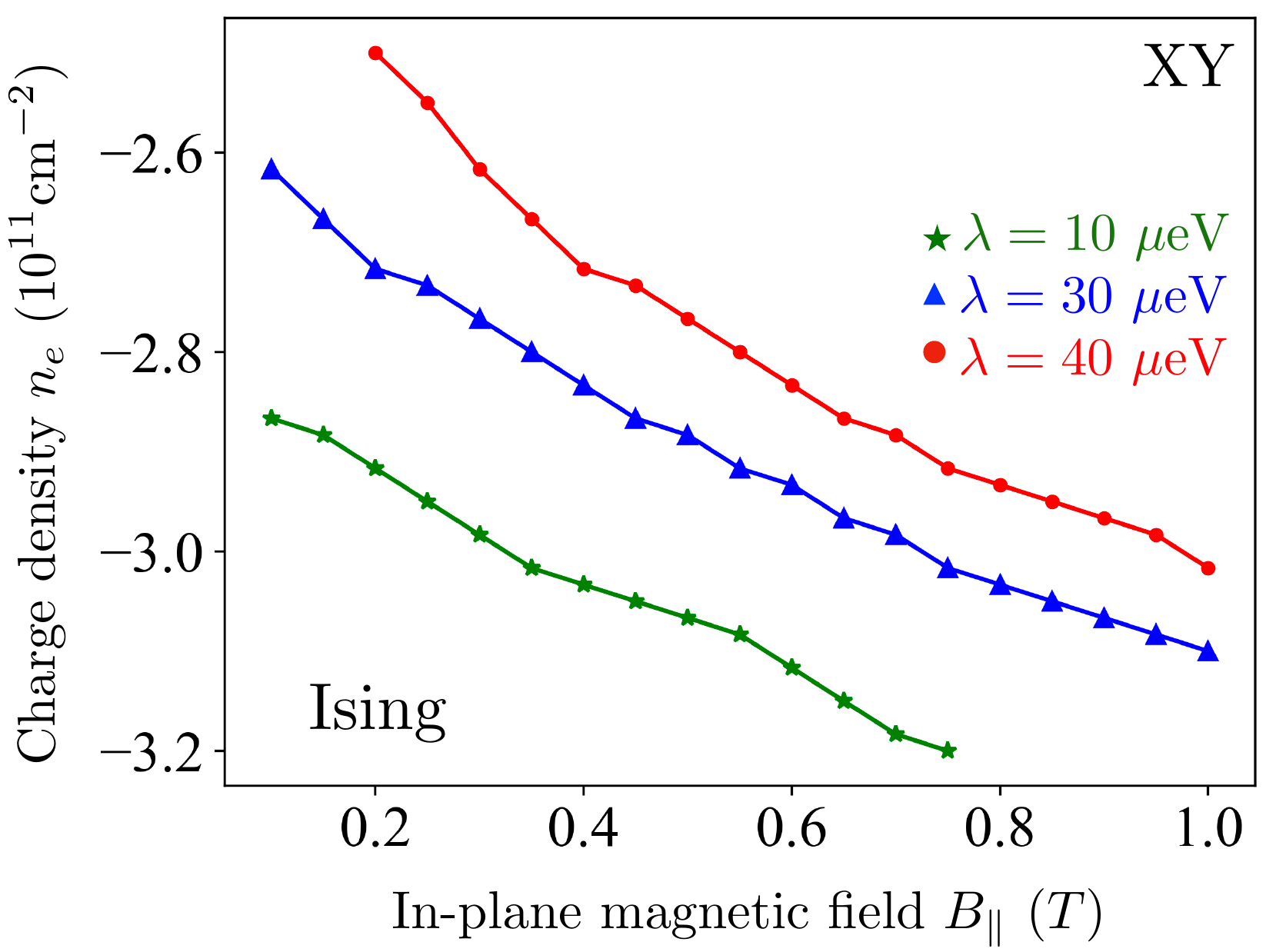}
    \caption{The displacement of the valley-Ising and valley-XY phase-boundary induced by in-plane magnetic field depends on the strength of spin-orbit coupling $\lambda$. Experimental observation of these shifts can be used to estimate $\lambda$.}
    \label{fig:soc_1}
\end{figure}


In this article, we present a comprehensive study of quarter metal phase diagram in the parameter space spanned by $n_e, U$, in-plane magnetic field $B_\parallel$ and SOC in rhombohedral trilayer graphene (RTG). Our findings indicate two distinct magnetic ground states for the quarter metal, termed the valley-XY phase and valley-Ising phase, separated by a first order phase transition. The momentum distribution of their holes either forms a conventional Fermi sea or adopts an annular shape, with the two types being separated by an annular Lifshitz transition (ALT).
This ALT corresponds to the enhanced resistance (depicted in yellow) line running from $(n_e,D)\sim(-4.5\times 10^{11} \text{cm}^{-2},0.35\text{V/nm})$ to $(n_e,D)\sim(-4.2\times 10^{11} \text{cm}^{-2},0.3\text{V/nm})$ in Fig.~1.e of Ref.~\cite{zhou_superconductivity_2021}.
The microscopic origin of magnetic anisotropic energy is attributed to the non-identical nature of wavefunctions across opposite valleys, SOC and $U$. Notably, the Ising-XY phase-boundary shift generated by an in-plane field $B_{\parallel}$ is strongly influenced by the strength of Kane-Mele spin-orbit coupling $\lambda$.
This correlation enables us to relate the slope $dn_e/dB_{\parallel}$ to the strength of the spin-orbit coupling, as shown in Fig.~\ref{fig:soc_6}. Additionally, we have calculated the in-plane magnetic susceptibility within the valley-Ising phase. These theoretical outcomes can be readily corroborated with experimental data to provide deeper insights into  graphene magnetism.

The remainder of this article is structured as follows: In Sec.~II, we study the quarter metal phase diagram without SOC. Sec.~III discuss the role of layer-polarization in assisting the  reentrance of paramagnet state at low densities. Sec.~IV examines the impact of SOC on the magnetic states within the quarter metal. In Sec.~V, we use the displacement of the Ising-XY phase boundary to estimate the magnitude of SOC. In Sec.~VI, we compare the quarter metal phase diagram between Bernal bilayer graphene and rhombohedral trilayer graphene. Finally, we summarize our work in Sec.~VII.

\section{Quarter-Metal Phase Diagram without Spin-Orbit Coupling}

In this section, we study the evolution of the quarter-metal groundstate in the $n_e-U$ parameter space when spin-orbit coupling is neglected. We first discuss the bandstructure of the valley-Ising phase $\ket{\theta_v=0}$ and the valley-XY phase $\ket{\theta_v=\pi/2}$ obtained from self-consistent mean-field calculation. Then we compare their band-energy and Fock exchange-energy evolution as a function of $n_e$. These analysis reveal that the first-order phase transition between $\ket{\theta_v=\pi/2}$  and $\ket{\theta_v=0}$ is driven by the decrease of exchange energy associated with the winding of pseudospin enforced by the band topology of multilayer graphene.

In the absence of spin-orbit coupling, the spin-quantization axis of a spin-polarized groundstate can be rotated without any energy cost. In contrast, there is a preferred direction to polarize valley degree of freedom because the wavefunctions and energy dispersion are different in the two valleys.

We use self-consistent Fock approximation to estimate the groundstate energy of the quarter metal state. The itinerant exchange effect, as described by Herring \cite{herring1966magnetism}, is arguably the most important factor in shaping the total energy of a quarter metal. This assertion is supported by the oscillation data, which strongly suggest that time-reversal and/or inversion symmetry of the quarter metal is broken to maximize its pseudospin polarization.
Given these observations, it seems likely that the self-consistent mean-field approximation would work better for the quarter metal than for other identified magnetic states, such as the half-metal. The mean field eigenvalue equation is $\hat{H}_{\vec{k}}|\psi_{n\vec{k}}\rangle=\epsilon_{n\vec{k}}|\psi_{n\vec{k}}\rangle$ where 
\begin{equation} \label{eq:H_MF}
    \hat{H}_{\vec{k}} = \hat{T}_{\vec{k}} + \hat{\Sigma}^{F}_{\vec{k}}.
\end{equation}
The band-Hamiltonian $\hat{T}_{\vec{k}}$ is parameterized by the experimentally informed \cite{zhou_half_2021} Slonczewski-Weiss-McClure parameters and the Fock self-energy is given by $\hat{\Sigma}^{F}_{\vec{k}}=-\sum_{n\vec{k}'}V_{\vec{k}-\vec{k}'}    \hat{\rho}_{\vec{k}'}$.
Here $V_{\mathbf{q}}$ is the Fourier components of the gate-screened Coulomb potential $V_{\mathbf{q}}=\frac{2\pi k_e}{\mathcal{A}}
\tanh(|\mathbf{q}|d)/(\epsilon_r |\mathbf{q}|)$ where $\mathcal{A}$ is area of the sample, Coulomb constant $k_e=1.44$ eVnm, screening constant $\epsilon_r=15$ and $d=5$ nm is the distance from the gate to the material. The density-matrix is constructed from the eigenvector and eigenvalue of the self-consistent Hamiltonian:
\begin{equation}
   \hat{\rho}_{\vec{k}}= \sum_{n} n_{F}(\epsilon_{n\vec{k}})|\psi_{n\vec{k}}\rangle\langle \psi_{n\vec{k}}|.
\end{equation}
The total energy of the self-consistent mean-field approximation is a sum of the band-energy $E_{b}^{\theta_v}$ and the Fock exchange energy $E_{ex}^{\theta_v}$,
\begin{align} \label{eq:E_band}
    E_{b}^{\theta_v}=\int \frac{d^2k}{4\pi^2} \text{Tr}[\hat{T}_{\vec{k}}\hat{\rho}_{\vec{k}}^{\theta_v}],
\end{align}
\begin{align} \label{eq:E_fock}
    E_{ex}^{\theta_v}=\frac{1}{2}\int \frac{d^2k}{4 \pi^2} \text{Tr}[\hat{\Sigma}_{\vec{k}}^F \hat{\rho}_{\vec{k}}^{\theta_v}],
\end{align}
where the superscript $\theta_v$ stands for the valley-polarization of the converged density matrix:
\begin{equation}
   \int \frac{d^2k}{4\pi^2}\text{Tr}[\hat{\rho}_{\vec{k}}^{\theta_v} (\tau_x\hat{\bf{x}}+\tau_z\hat{\bf{z}})]
    \sim \sin(\theta_v)\hat{\bf{x}}+\cos(\theta_v)\hat{\bf{z}}.
\end{equation}
Here the trace $\text{Tr}$ operates on the $24$ dimensional space generated by the 6$p_z$ orbitals, the $2$ spin and $2$ valley degree of freedoms. More discussion of the Hamiltonian can be found in Ref.~\cite{huang2023spin} and Ref.~\cite{das2023unconventional}. We used a Lagrange multipliers to generate the energy landscape associated with order parameter rotations.


\begin{figure}[t]
    \centering
    \includegraphics[width=\columnwidth]{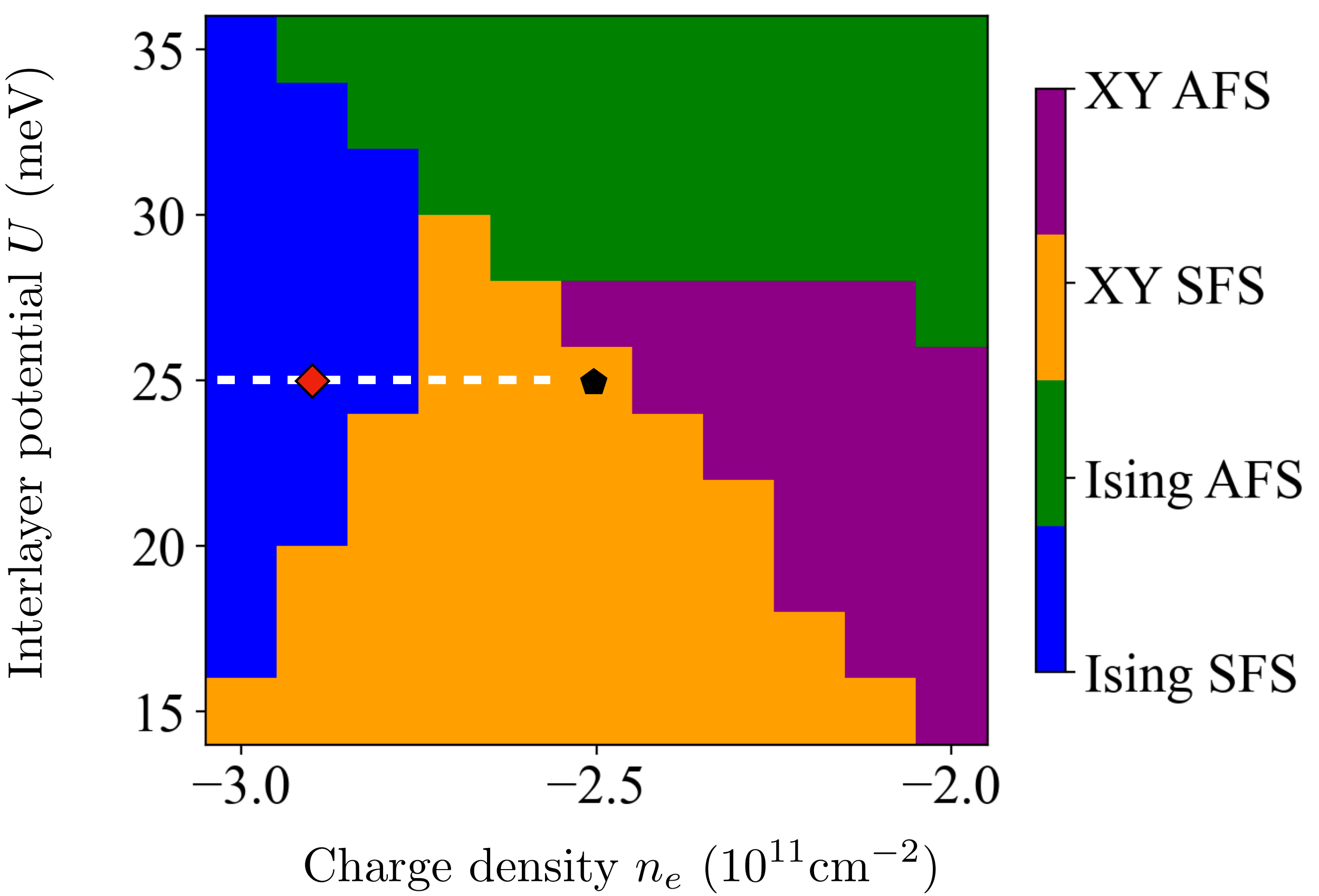}
    \caption{Mean field phase diagram shows 4 unique phases of quarter metal distinguished by their valley-polarization (XY/Ising) and Fermi surface topology (Simple Fermi Sea/Annular Fermi Sea). 
    The evolution of band and Fock energy along the white dashed line, as well as the bandstructure of the black and red points are further expanded in Fig.~\ref{fig:soc_3}.
    }
    \label{fig:soc_2}
\end{figure}
\begin{figure*}[t]
    \centering
    \includegraphics[width=2\columnwidth]{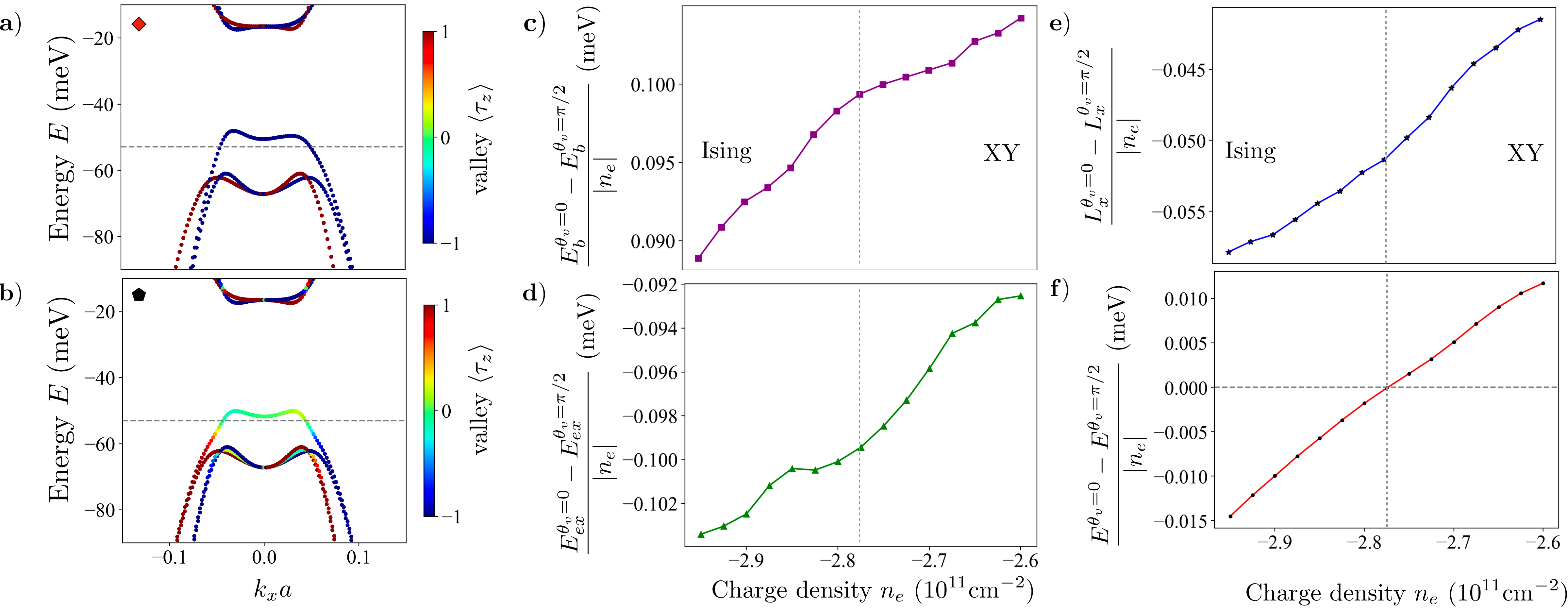}
    \caption{\textbf{a)} Band-structure  (along $k_ya=0$) of valley-Ising phase at $n_e=-2.9\times 10^{11}\text{cm}^{-2}$ and \textbf{b)} valley-XY phase at $n_e=-2.5\times 10^{11}\text{cm}^{-2}$ respectively at $U=25~\text{meV}$. The chemical potential $(\mu)$ is marked by gray horizontal dashed line in both cases. \textbf{c,d}) Band and Fock exchange energy difference of valley-Ising and valley-XY phase per charge as a function of density (marked by white dashed line in Fig.~\ref{fig:soc_2}) shows the energy difference is increasing while approaching charge neutrality. The band energy in Ising phase is higher than valley-XY phase, whereas exchange energy of valley-XY phase is higher than valley-Ising phase. \textbf{e}) Difference in layer pseudospin average $L^{\theta_v}_x$ per charge in valley-Ising and valley-XY state decreases while approaching charge neutrality. This indicates  why exchange energy difference between these two states are increasing with decreasing hole density. \textbf{f}) The total energy
    difference between valley-Ising and valley-XY phase $v.s.$ $n_e$ identifies the phase-transition at $n_e\approx -2.78\times 10^{11}$cm$^{-2}$.
    The positive slope indicates the chemical potential of the Ising state is higher than the XY state.
}
    \label{fig:soc_3}
\end{figure*}

As shown in Fig.~\ref{fig:soc_2}, 
we identified four unique phases of the quarter metal. These phases are characterized by differences in the number of Fermi surfaces and their valley-polarization. They are separated by first-order Ising-XY transition boundaries and first-order annular-Lifshitz transition phase boundaries. The simplest phase among these four appears at large $|n_e|$ and $U$. 
It has an Ising-like valley-polarization ($|\theta_v=0\rangle$ or $|\theta_v=\pi\rangle$)
and all holes are enclosed in a simply-connected Fermi surface (SFS). 
Designated as the Ising-SFS, this phase is colored blue in Fig.\ref{fig:soc_2}
Its unconventional relationship between  orbital magnetization and valley-polarization is discussed in Ref.~\cite{das2023unconventional}.
As $|n_e|$ decreases in the interlayer potential range $30<U<35$meV, the Ising-SFS undergoes an annular Lifshitz transition (ALT) where an electron-like Fermi surface appears at the centre of the Fermi sea. This leads to an expansion of the hole-like Fermi surface without changing the valley/spin polarization. We refer to this state as the Ising-AFS which is represented as green in Fig.~\ref{fig:soc_2}. In contrast, when $|n_e|$ decreases at smaller interlayer potential $17<U<30$meV, we found that the Ising SFS goes through a magnetic phase transition and changes its valley order parameter from $\theta_v=0$ to $\theta_v=\pi/2$. We dubbed this state as XY-SFS because it enclosed all the holes in a single Fermi surface. Note that the Fermi surface is intrinsically tied to the valley-order parameter: it adopts a $C_6$ symmetry for $\theta_v=\pi/2 $ and $C_3$ for $\theta_v=0$. As $|n_e|$ reduces to smaller values, XY-SFS also encounters an ALT. We've labelled this phase as XY-AFS, and it is distinguished by purple color in Fig.~\ref{fig:soc_2}.

Fig.~\ref{fig:soc_3}.\textbf{a}) and \textbf{b}) shows the bandstructure $\epsilon_{n,kx,ky=0}$ $v.s.$ $k_x$ for the valley-Ising and valley-XY states respectively. 
While the low-energy electronic states in multilayer graphene are centered at the zone-corner, we fold the bands into the $\Gamma$ point to visualize valley-mixing \cite{PhysRevLett.126.056801}.
The annular Lifshitz transition appears because the bandstructure has a local minimum at $\vec{k}=0$, as shown in Fig.~\ref{fig:soc_3}\textbf{b}). 
The valence band-projected valley-polarization shows that the quasiparticles in valley-XY states is a linear superposition of states from opposite valleys. Such valley-mixing is enabled by tripling of the area of the real-space unit cell. In contrast, the valley Ising state merely redistribute the holes all into one valley (say $\tau_z=-1$ in Fig.~\ref{fig:soc_3}.\textbf{a}) without changing the valley content of the spinor.

In order to understand the Ising-XY phase transition, we break down the energy difference between $\ket{\theta_{v}=0}$ and $\ket{\theta_{v}=\pi/2}$ into the band contribution and exchange contribution across various $n_e$. As illustrated by the purple curve in Fig.~\ref{fig:soc_3}.\textbf{c}), the band-energy of the Ising state is higher than that of the XY state. 
This is because the $XY$ state requires a lesser degree of quasiparticle redistribution from the paramagnetic state. 
Consequently, from a band energy perspective, the XY state is favored over the Ising state as the quarter metal's magnetic ground state. Importantly, this preference is reinforced at lower $|n_e|$ thus the band energy contributes to driving the density-induced first-order Ising-XY phase transition. The chemical potential difference between Ising and XY state from the band-energy can be inferred from the slope in Fig.~\ref{fig:soc_3}.\textbf{c}), 
\begin{equation} \label{eq:mu_diff_band}
   \frac{\partial(E_{b}^{\theta_v=0}-E_{b}^{\theta_v=\pi/2})}{\partial n_e} \equiv \mu_{b}^{\theta_v=0}-\mu_{b}^{\theta_v=\pi/2}>0.
\end{equation}

\begin{figure*}[t]
    \centering
    \includegraphics[width=2\columnwidth]{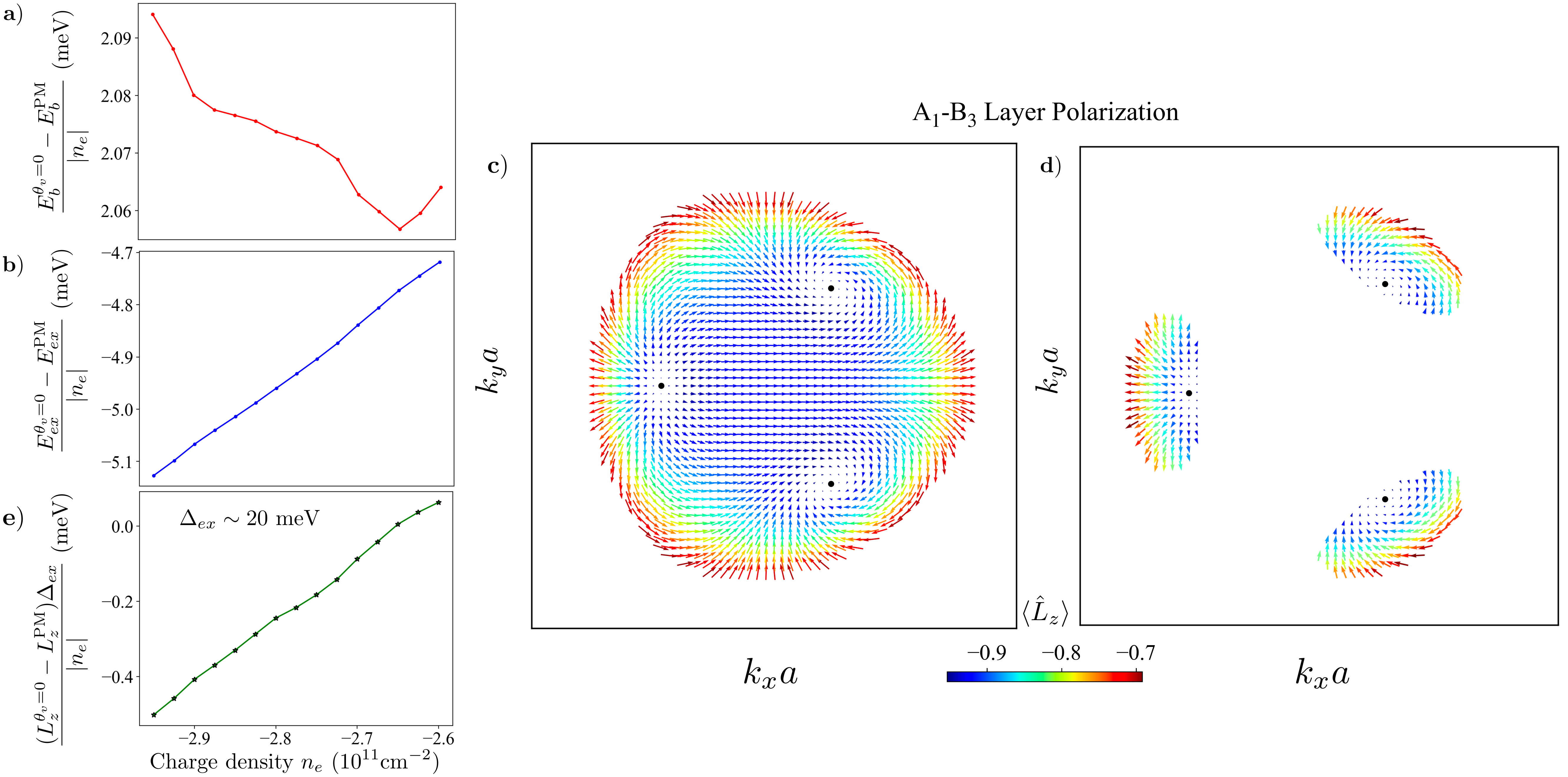}
    \caption{\textbf{a}) The band energy of the Ising state with respect to paramagnetic state decreases while approaching charge neutrality.  \textbf{b}) The Ising Fock energy increases linearly with respect to paramagnetic Fock energy as $|n_e|$ deceases. 
    \textbf{c}) Layer pseudospin orientation of the holes enclosed in a $C_3$ Fermi surface in the valley-Ising quarter metal. As we circle the Fermi surface anticlockwise, the pseudospin rotates by $6\pi$. 
    Result is obtained at $n_e,U=-2.9\times 10^{11}\text{cm}^{-2},25~\text{meV}$.
    \textbf{d}) Layer pseudospin orientation of the hole-like Fermi pockets shown in one valley in paramagnetic phase at same $n_e,U$. \textbf{e}) The layer pseudospin $L_z$ of paramagnetic phase changes faster than valley-Ising phase with respect to density, and consequently the exchange energy gap between valley-Ising and paramagnetic phase shrinks towards charge neutrality.
    }
    \label{fig:soc_4}
\end{figure*}

Fig.~\ref{fig:soc_3}.\textbf{d}) shows that the XY state's exchange-energy is elevated compared to the Ising state's exchange-energy. This energy difference can be traced back to the non-uniform alignment of quasiparticle valley-polarization in the XY state, as shown by the color variations in Fig.\ref{fig:soc_3}.\textbf{a}). However, as $|n_e|$ decreases, the difference between these exchange energies $E_{ex}^{\theta_v=\pi/2}-E_{ex}^{\theta_v=0}$ narrows, leading to
\begin{equation}\label{eq:mu_diff_ex}
   \frac{\partial(E_{ex}^{\theta_v=0}-E_{ex}^{\theta_v=\pi/2})}{\partial n_e} \equiv \mu_{ex}^{\theta_v=0}-\mu_{ex}^{\theta_v=\pi/2}>0.
\end{equation}
This trend is attributed to the reduction of interlayer coherence $ L_{x}^{\theta_v=\pi/2}-L_{x}^{\theta_v=0}$ with decreasing $|n_e|$, as shown in Fig.~\ref{fig:soc_3}\textbf{e}). The interlayer coherence and layer-polarization (per area) is defined as $L_j^{\theta_v} = \int \frac{d^2k}{4\pi^2}L_j^{\theta_v}(\vec{k})$ where $j=x,y,z$:
\begin{align} \label{eq:layer}
    L_x^{\theta_v}(\vec{k})&=\text{Tr}[\hat{\rho}_{\vec{k}}^{\theta_v} (\ket{A1}\bra{B3}+\ket{B3}\bra{A1})],\\
    L_y^{\theta_v}(\vec{k})&=-i\text{Tr}[\hat{\rho}_{\vec{k}}^{\theta_v} (\ket{A1}\bra{B3}-\ket{B3}\bra{A1})],\\
    L_z^{\theta_v}(\vec{k})&=\text{Tr}[\hat{\rho}_{\vec{k}}^{\theta_v} (\ket{A1}\bra{A1}-\ket{B3}\bra{B3})].
\end{align}
Here $\ket{A1}$ and $\ket{B3}$ represent the outermost layer sublattices. Note the antisymmetric layer coherence remains zero for both type of valley-polarization $L_y^{\theta_v=0}=
L_y^{\theta_v=\pi/2}=0$. For all $n_e$ and $U$, although the layer-polarization is greater in magnitude than layer coherence, for example :~$L_z^{\theta_v=0}/|n_e|\sim -7.584, L_z^{\theta_v=\pi/2}/|n_e|\sim-7.583$ and $L_x^{\theta_v=0}/|n_e|\sim-1.283,~
L_x^{\theta_v=\pi/2}/|n_e|\sim-1.228$ at $n_e,U=-2.9\times 10^{11}~\text{cm}^{-2},25~\text{meV}$, their difference  
$(L_{z}^{\theta_v=\pi/2}-L_{z}^{\theta_v=0})/|n_e|\sim 10^{-3}$ is an order of magnitude smaller than those shown in Fig.~\ref{fig:soc_3}.\textbf{e}). 
This is because the $L_{z}$ is mainly determined by the interlayer potential and the position of the Fermi level.



As a result, while the exchange energy leans towards the Ising state as the magnetic ground state of quarter metal, this inclination becomes less pronounced at smaller $|n_e|$.
In this sense, the exchange energy also contributes to driving the Ising-XY phase transition as $|n_e|$ reduces. Fig.~\ref{fig:soc_3}.\textbf{f}) plots the total energy difference between Ising and XY and indicated the phase-boundary at $n_e\sim-2.78\times10^{11}$cm$^{-2}$.

\section{Re-entrance of Paramagnetic Metal at Low Density and the Layer Vorticity}

In the large interlayer potential ($U$) and lowest-density ($n_e$) region of the $n_e-U$ phase-diagram for multilayer graphene, an interesting question arises: Why does the spin-and-valley polarized magnetic quarter-metal transition into a paramagnetic metal as density ($|n_e|$) decreases, in sharp contrast to the two-dimensional electron gas (2DEG)? In this section, we discuss how the distinctive topological band properties of multilayer graphene leads to a narrowing energy gap between the magnetic states and paramagnetic state as $|n_e|$ decreases. This trend provides a crucial ingredient in understanding the re-entrance of the paramagnetic phase observed at low densities \cite{zhou_half_2021,zhou_isospin_2021}.
 
%

Fig.~\ref{fig:soc_4}.\textbf{a}) and \textbf{b}) plot the band-energy and exchange-energy per particle of the Ising and XY states relative to the paramagnetic state. The first feature to notice is that the energy difference of the Ising and XY magnetic states relative to the paramagnetic state significantly surpasses the energy gap between the Ising and XY states themselves, i.e., the magnetic anisotropic energy (c.f.~Fig.~\ref{fig:soc_3}.\textbf{c},\textbf{d} and Fig.~\ref{fig:soc_4}.\textbf{a},\textbf{b}).
A second important feature is the swift decline in exchange energy per particle of the magnetic state relative to the paramagnetic state as $|n_e|$ decreases. Meanwhile, the band-energy difference per particle stays relatively constant. This disparity, which is on the order of a magnitude, contrasts significantly with behaviors seen in 2DEG.
In 2DEG, as $n_e$ decreases, the kinetic energy difference between the fully spin-polarized state and the paramagnetic state, 
\begin{equation}
x    (E_{b}^{FM}-E_{b}^{PM})/n_e= \frac{1}{r_s^2}\text{Ry} 
\end{equation}
changes more quickly than their Fock energy difference 
\begin{equation}
    (E_{ex}^{FM}-E_{ex}^{PM})/n_e=-\frac{8\sqrt{2}}{3\pi}\frac{(\sqrt{2}-1)}{r_s}\text{Ry},
\end{equation} 
where $\pi (r_s a_B)^2=\frac{1}{n_e}$ and $\text{Ry}=13.6$eV.
We found that such different density-dependence of the band and Fock energies between 2DEG and multilayer graphene arises from the topological band properties of multilayer graphene.
In a $N$-layer rhombohedral graphene electronic system, the first valence and conduction band can be approximately described by an effective two-orbital Hamiltonian comprised of the outermost layer $p_z$ orbitals, e.g.~$A_1-B_N$. These two degrees of freedom $A_1$ and $B_N$ form a layer pseudospin and it winds $2\pi N$ around the Brillouin zone corner, leading to a vortex structure in the momentum space. The exchange energy, which favors a uniform distribution of layer pseudospin in momentum space challenges this vortex configuration. See \cite{PhysRevB.77.041407} for more details.

Fig.~\ref{fig:soc_4}.\textbf{d}) and \textbf{e}) depict the valence-band projected layer pseudospin orientation inside the hole-like Fermi surface for both the valley-Ising and paramagnetic phases. Arrows indicate the in-plane layer-polarization $L_x(\vec{k}),L_y(\vec{k})$ and (c.f Eq.~\eqref{eq:layer}) the color indicates the values of $L_z(\vec{k})$. These vortex cores are highlighted by the black dots. The vortex cores are situated close to the valence band maximum and their corresponding layer polarization $(L_x,L_y,L_z)\sim (0,0,-1)$.  As the pseudospins approach the vortex cores (or the band-edge), they changes rapidly to align their pseudospin towards the $(0,0,-1)$ directions.

In the paramagnetic state, the Fermi level is situated very close to the valence band edge with a Fermi energy of $\epsilon_F\sim 2.5~\text{meV}$ at $n_e,U=-2.9\times 10^{11}~\text{cm}^{-2},25~\text{meV}$.
Thus, $L_z^{PM}$ changes rapidly with decreasing $|n_e|$.  On the other hand, in the quarter metal (be it Ising or XY), the area of the single hole-like Fermi surface expands by a factor of 12 compared to the area enclosed by the small hole-like Fermi pocket in the paramagnetic state. This causes the Fermi energy
to be situated further from the valence band edge even when both states share the same values of $n_e$ and $U$, with $\epsilon_F\sim 4.8$ meV.
Therefore, $L_z^{\theta_v=0,\pi/2}$ changes slowly with decreasing $|n_e|$. In Fig.~4e), we show $(L_z^{\theta_v=0}-L_z^{PM})\Delta_{ex}/|n_e|$ $v.s.$ $n_e$ and note that this slope closely resembles the slope in $(E_{ex}^{\theta_v=0}-E_{ex}^{PM})/|n_e|$ $v.s.$ $n_e$ shown in Fig.~4b). 
Thus, this density-dependence of the layer-polarization can reduce the exchange energy difference between the quarter metal and paramagnetic state thereby significantly influencing the re-emergence of the paramagnetic state.

\begin{figure*}[t]
    \centering
    \includegraphics[width=2.0\columnwidth]{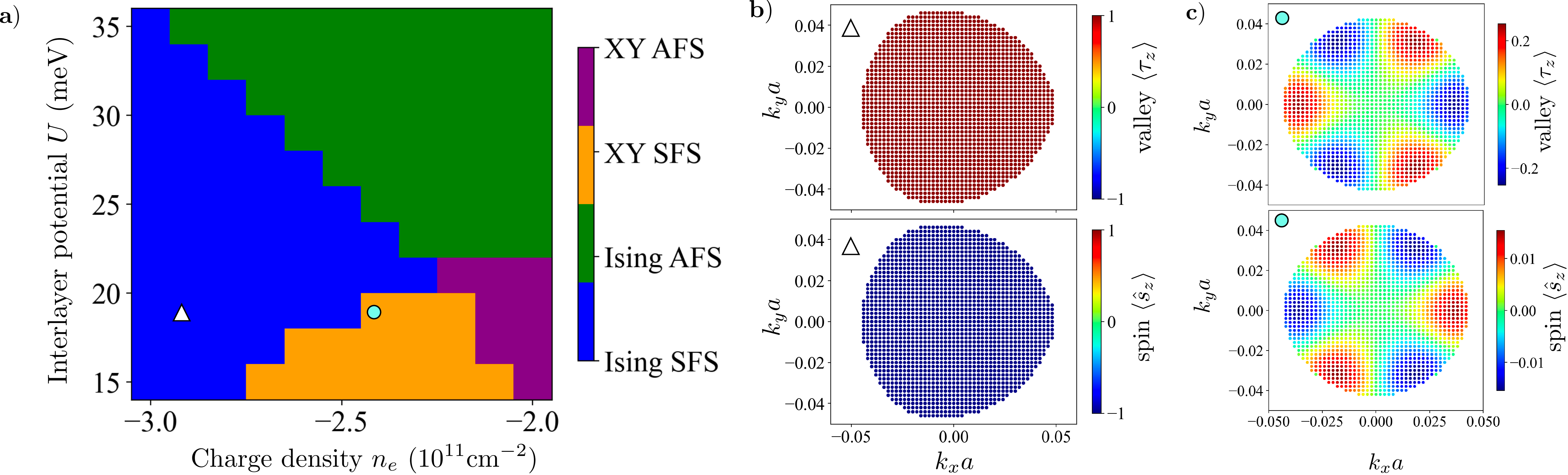}
    \caption{\textbf{a}) Modification of the phase diagram in Fig.~\ref{fig:soc_2} due to a Ising spin-orbit coupling of $\lambda=30~\mu$eV. \textbf{b}) valley and spin texture of valley-Ising state with simply connected Fermi surface (Ising-SFS) for $n_e,U=-2.9\times 10^{11}~\text{cm}^{-2},18~\text{meV}$.\textbf{c}) valley and spin texture of valley-XY state with simply connected Fermi surface (XY-SFS) for $n_e,U=-2.4\times 10^{11}~\text{cm}^{-2},18~\text{meV}$. }
    \label{fig:soc_5}
\end{figure*}

\section{Effect of Spin-Orbit Coupling On Quarter-Metal}

It is not unreasonable to assume that SOC plays a more important role in shaping the magnetic anisotropy landscape compare to the lattice-scale Coulomb interaction \cite{PhysRevB.85.155439}\footnote{in preparation, Guopeng Chunli}. 
When we approximate lattice-scale Coulomb interaction by a delta-function interaction, it is clear that it won't alter the distribution of particles with identical spin-valley quantization. 
Notably, unlike the zeroth-Landau level wavefunctions in graphene where sublattice and valley degrees of freedom are identical, the quasiparticle wavefunctions in RTG exhibit rather complex orbital dependence.
Nevertheless, a direct mean-field calculation suggests that inter-valley scattering with a magnitude of $u_\perp=0.5~\text{meV}$ has very small impact on the quarter-metal phase diagram [see Appendix A]. Therefore, our analysis of magnetic anisotropy energy of quarter metal will center on the SOC.

Since spin-orbit coupling that is off-diagonal in layer degree of freedom is suppressed by the application of large electric displacement field, we focus on the simple momentum independent Kane-Mele-like SOC,
\begin{equation} \label{eq:SOC}
    \hat{H}_{KM} = \lambda ~\hat{\sigma}_z\hat{\tau}_z\hat{s}_z.
\end{equation}
This Hamiltonian reduces the full $SU(2)$ spin-rotation symmetry down to $U(1)$ symmetry as it only commutes with the generator of $s_z$ rotation: $[\hat{H}_{KM},e^{i\phi_s \hat{s}_z}]=0$.  Note $\hat{\sigma}_z=\ket{A1}\bra{A1}-\ket{B3}\bra{B3}$ here is taken as the layer polarization between the $A1-B3$ sublattice \cite{mccann2013electronic,PhysRevB.77.041407}. This form of SOC can be derived by considering the point group symmetry of rhombohedral trilayer graphene \cite{PhysRevB.95.165415,PhysRevB.82.245412}.

When adding Eq.~\ref{eq:SOC} (with $\lambda=30~\mu$eV) to Eq.~\ref{eq:H_MF}, and carry out the self-consistent calculations, we obtained the phase diagram shown in Fig.\ref{fig:soc_5}.\textbf{a}). A quick comparison between Fig.~\ref{fig:soc_5}.\textbf{a}) and Fig.~\ref{fig:soc_2} shows that even a modest value of SOC ($\lambda=30~\mu$eV) has a pronounced effect on the $n_e-U$ phase diagram for the quarter metal. Note $\lambda$ has almost no effect on the annular Lifshitz transition (ALT) phase boundaries. This is because the two phases on both side of the ALT phase boundary have identical spin-valley order parameters. Thus, SOC lowers their energy in a similar way. In contrast, the Ising-XY phase boundary is shifted dramatically in the direction that expands the Ising phase, strongly influencing the magnetic anisotropy energy landscape.  

The principal effect of $\hat{H}_{KM}$ is to introduce correlation between the quasiparticle valley-polarization and its spin-quantization. In order to understand this correlation, we plot the momentum space distribution of spin and valley-polarization for the holes in the Ising and XY state. 
For the Ising ground state, the spin polarization of all the holes are aligned in the $+z$ direction while their valley-polarization are all aligned in the $-z$ direction. This state is represented by the notation $|\theta_v,\theta_s\rangle$, specifying the valley-polarization angle $\theta_v$ and the spin polarization angle $\theta_s$. This state is called the Ising state because the energies of $|\theta_v=\pi,\theta_s=0\rangle$ and its time-reversed counterpart $|\theta_v=0,\theta_s=\pi\rangle$ are the same. The effect of SOC in the Ising phase is to introduce an uniaxial easy-axis normal to two-dimensional material plane. Note while the spin-quantization axis contains spatial information (in this case, it's perpendicular to the two-dimensional material plane), the valley-polarization, lacks any spatial interpretation.
In addition to the the Ising-state, we identified another stationary solution characterize an order parameter defined by non-zero values of  $\tau_x$, $s_x$ and $s_z\tau_z$. We named this stationary solution the vallley-XY, denoted as 
$\ket{\theta_s=\pi/2,\theta_v=\pi/2}$. Note the total energy of this state remains unchanged under rotations generated by  $e^{i\phi_s s_z}$ and  $e^{i\phi_v \tau_z}$ rotations, justifying the valley-XY nomenclature. 
Fig.~\ref{fig:soc_5}.\textbf{c}) shows the valley-polarization and spin-polarization of $\ket{\theta_s=\pi/2,\theta_v=\pi/2}$ for the holes enclosed inside the $C_6$ Fermi surface. Navigating around the Fermi surface clockwise, the valley-polarization exhibits an oscillatory pattern about the $\tau_x$ axis. This specific valley-polarization configuration aims to minimize both the exchange and band energy: it minimizes the exchange energy by predominantly aligning the valley-polarization along the x-direction and introduces a valley-population imbalance across six unique momentum regions to further reduce band energy.
The presence of a finite SOC generates a similar oscillatory pattern in spin so that the valley-polarization ($\bra{\psi_{nk}} \hat{\vec{\tau}} \ket{\psi_{nk}}$) is almost locally anti-parallel to spin-polarization ($\bra{\psi_{nk}} \hat{\vec{s}} \ket{\psi_{nk}}$) in momentum space. Contrary to the Ising state, which reduces its energy linearly with SOC, the XY state decrease its energy quadratically in $\lambda$.

\begin{figure}[t]
    \centering
    \includegraphics[width=\columnwidth]{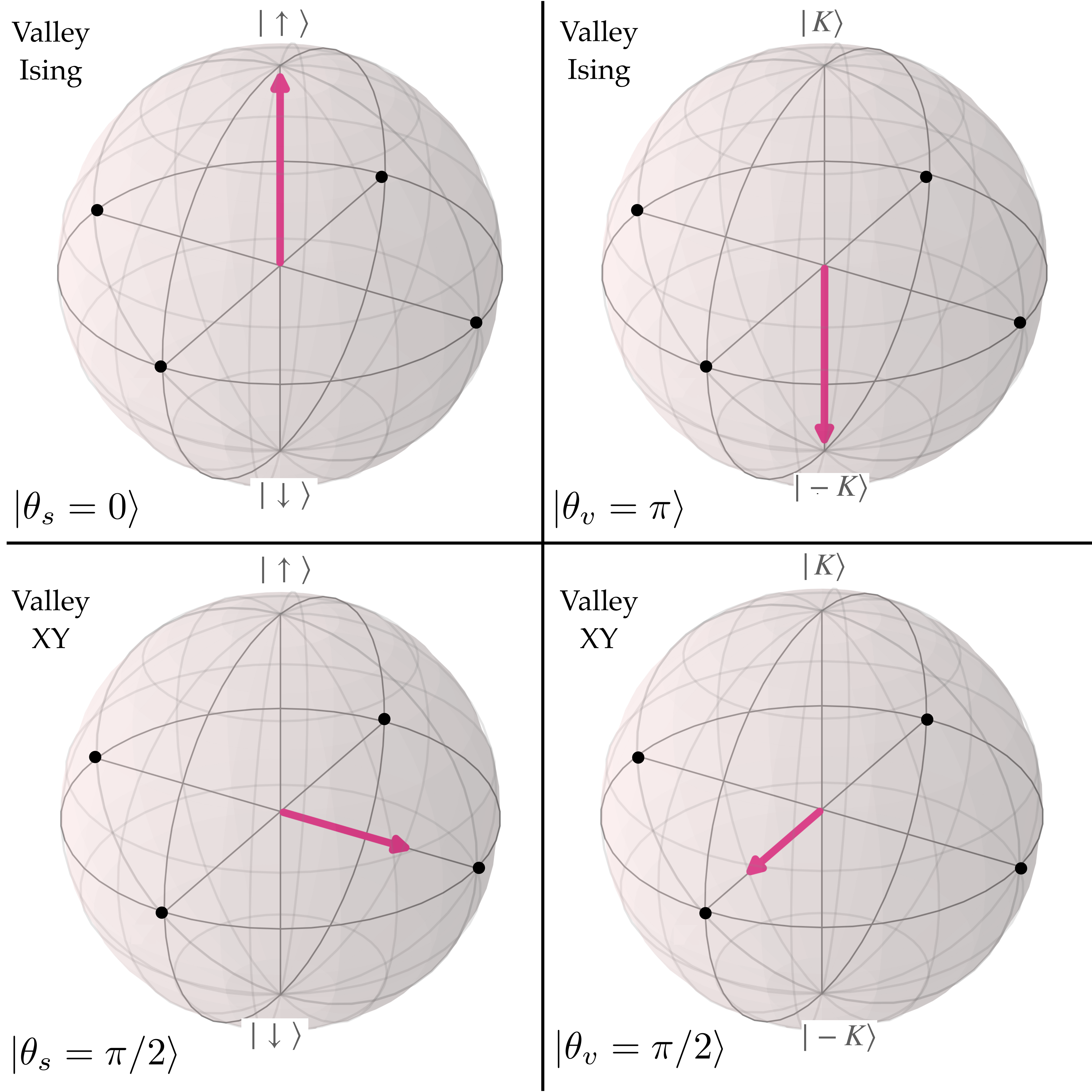}
    \caption{
The effect of Kane-Mele spin-orbit coupling (SOC) on the order parameters of valley-Ising and valley-XY quarter metals is illustrated on the two Bloch sphere. The $\pm z $ directions on the these spheres correspond to the $|\uparrow,\downarrow\rangle$ and $|K,K'\rangle$ degrees of freedom. In the valley-Ising state, SOC pins the spin and valley polarization in the $\pm z$  and $\mp z$ directions. Conversely, in the valley-XY phase, it aligns them in directions orthogonal to $z$. The XY-phase features two soft-modes, arising from independent rotations of spin and valley order parameter within the magnetic easy-plane. Refer to the maintext (Sec.~IV) for discussion on the magnitude of the arrows.
    }
    \label{fig:spin_valley_bloch}
\end{figure}

The valley-XY and valley-Ising order parameters are visualized on two Bloch spheres in Fig.~\ref{fig:spin_valley_bloch}. Arrows touching the Bloch sphere represent maximium polarization, as seen in the valley-Ising state. Note, the valley-XY phase lacks maximum spin and valley polarization because of its non-uniform momentum distribution of spin-polarization and valley-polarization as shown in Fig.~\ref{fig:soc_5}.\textbf{c}).

In the Appendix C, we use the Lagrange multiplier $\hat{H}_L(\vec{n})=\lambda_L[\vec{n}\cdot \hat{\vec{s}}+\vec{n}\cdot(e^{i\frac{\pi}{2}\hat{\tau}_x}\hat{\vec{\tau}}e^{-i\frac{\pi}{2}\hat{\tau}_x})]$
to further expand on the energy landscape of other possible quarter metal phases in the presence of spin-orbit coupling.

\section{Displacement of the Ising-XY phase boundary}

\begin{figure*}[t]
    \centering
    \includegraphics[width=1.9\columnwidth]{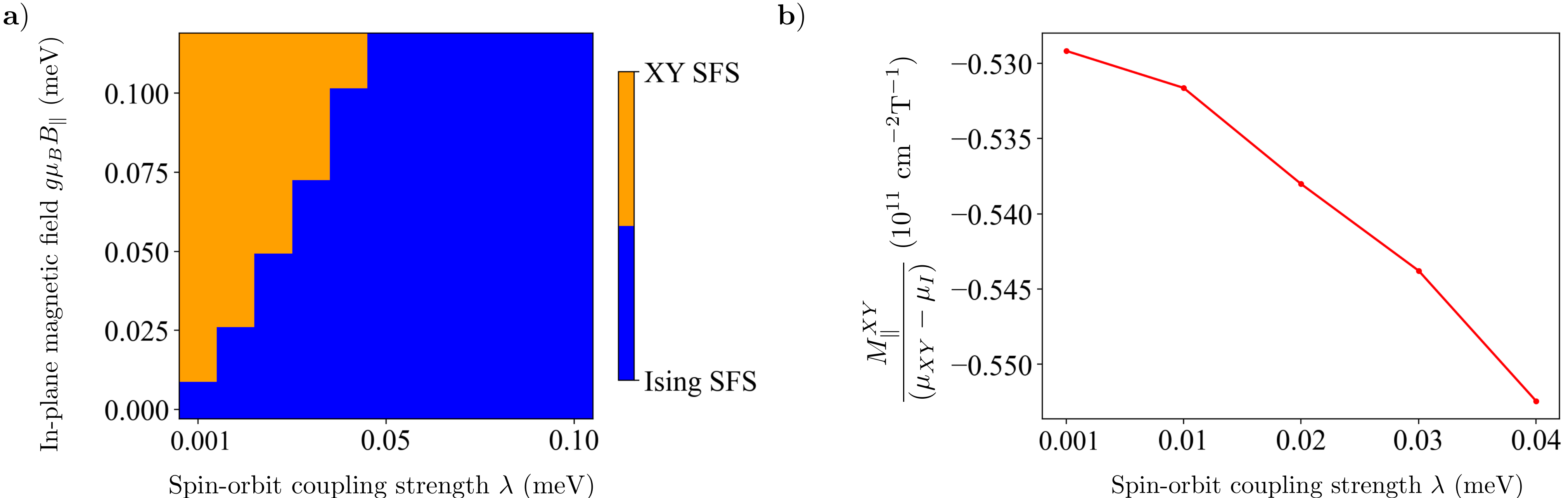}
    \caption{\textbf{a}) The slope of the phase boundary for $n_e,U=-2.9\times 10^{11}~\text{cm}^{-2},18~\text{meV}$ indicates that the rate at which $B_\parallel$ lowers the energy of valley XY state is less than the rate at which $\lambda$ lowers the energy of the valley Ising state. \textbf{b}) $\frac{ M^{XY}_\parallel}{(\mu_{XY}-\mu_I)}$ is calculated along the phase boundary which is equal to the slope $\frac{\partial n_e}{\partial B_\parallel}$ indicated in Fig.~\ref{fig:soc_1} for the specific $n_e$ and $U$.}
    \label{fig:soc_6}
\end{figure*}

In this section, we extend the electronic phase diagram of the quarter metal from the two-dimensional $n_e-U$ parameter space to the more comprehensive four dimensional $n_e-U-\lambda-B_{\parallel}$ space. Here, $\lambda$ denotes the Kane-Mele SOC (introduced in the last section) and $B_\parallel$ represents the in-plane Zeeman field. In this expanded parameter landscape, we first examine the movement of Ising-XY phase-boundary as a function of $\lambda$ and $B_\parallel$, keeping $n_e,U$ fixed. Then, we study the movement of this phase-boundary as a function of $n_e$ and $B_\parallel$, keeping $\lambda,U$ fixed. The latter analysis enables us to correlate  $\lambda$ to the experimentally observable metric $dn_e/dB_{\parallel}$.


We compute the $n_e-U-\lambda-B_{\parallel}$ phase diagram by solving the following self-consistent mean-field equation at various density $n_e$ and interlayer potential $U$:
\begin{equation} \label{eq:H_KM_B}
    \mathcal{H}_{\vec{k}} = \mathcal{T}_{\vec{k}}+ \Sigma_{\vec{k}}^F\;\; , \;\; 
    \mathcal{T}_{\vec{k}} =\hat{T}_{\vec{k}}+\hat{H}_{KM}+\hat{H}_{B_{\parallel}},
\end{equation}
where $\hat{T}_{\vec{k}}$ and $\Sigma_{\vec{k}}^F$ are described in Eq.~\ref{eq:H_MF}. The Kane-Mele SOC, $\hat{H}_{KM}$ is defined in Eq.~\ref{eq:SOC} and the Zeeman energy is given by the following
\begin{equation} \label{eq:zeeman}
    \hat{H}_{B_{\parallel}} = -\frac{1}{2}g\mu_B\hat{s}_x B_\parallel.
\end{equation}
Here $g=2$ is the gyromagnetic ratio, $\mu_B$ is the Bohr magneton, and $\hat{s}_{x}$ is the 1st Pauli matrix.
 The valley Ising-XY phase boundary is defined by the following equation:
\begin{equation}
    E^{\theta_v=0}(n_e,U,\lambda,B_{\parallel})=E^{\theta_v=\pi/2}(n_e,U,\lambda,B_{\parallel}).
\end{equation}
where the total energy is
\begin{align} \label{eq:E_total}
    E^{\theta_v}(n_e,U,\lambda,B_{\parallel})\equiv\frac{1}{2}\int \frac{d^2k}{4\pi^2} \text{Tr}[(\hat{\mathcal{T}}_{\vec{k}}+\mathcal{H}_{\vec{k}})\hat{\rho}_{\vec{k}}^{\theta_v}].
\end{align}
Fig.~\ref{fig:soc_6}.\textbf{a}) shows the movement of the Ising-XY phase-boundary as a function of $B_\parallel$ and $\lambda$ for fixed $n_e$ and $U$. 
To leading order in $B_{\parallel}$, the total energy of the state $\ket{\theta_v=\pi/2,\theta_s=\pi/2}$ is lowered while $\ket{\theta_v=0,\theta_s=\pi}$ is unaffected. In contrast, 
to leading order in $\lambda$, the total energy of the state $\ket{\theta_v=0,\theta_s=\pi}$ is lowered while $\ket{\theta_v=\pi/2,\theta_s=\pi/2}$ is unaffected.
The slope $d(g\mu_B B_{\parallel})/d\lambda>1$ means SOC is more effective in reducing the energy of the Ising state compared to the energy reduction achieved by the Zeeman effect on the XY-state
$\ket{\theta_v=\pi/2,\theta_s=\pi/2}$. This is simply because the out-of-plane spin-polarization of the Ising state is greater than the net in-plane spin-polarization of the XY-state, as shown in Fig.~\ref{fig:soc_5}.\textbf{b}) and \textbf{c}). Note that when the Zeeman field reaches a sufficiently high value (i.e.~$g\mu_B B_{\parallel}\gtrapprox \lambda$), the Ising state manifests field-induced in-plane spin-polarization, such that $\theta_s=\pi-\delta$ and $\theta_v=\delta$, with $\delta \ll \pi$.

Next, we consider the Ising-XY phase-boundary as a function of $n_e$ and $B_{\parallel}$ for fixed $\lambda$ and $U$. 
Since $n_e$ and $B_{\parallel}$ are thermodynamical conjugate to chemical potential and magnetic-moments, a small variation of density $n_e$ around Ising-XY boundary at $n_e^*$ and $B_{\parallel}$ around $0$, results in the following energy changes:
\begin{align}
    E^{\theta_v=0}(n_e,U,\lambda,B_{\parallel})&=
    E^{*}+
    \mu^{\theta_v=0}  dn_e - M_{\parallel}^{\theta_v=0}dB_\parallel \\
    E^{\theta_v=0}(n_e,U,\lambda,B_{\parallel})&=
    E^{*}+
    \mu^{\theta_v=\pi/2} dn_e - M_{\parallel}^{\theta_v=\pi/2}dB_\parallel
\end{align}
where $E^{*}= E^{\theta_v=0}(n_e^*,U,\lambda,B_{\parallel}=0)=E^{\theta_v=\pi/2}(n_e^*,U,\lambda,B_{\parallel}=0)$ is the equilibrium free energy energy when $B_{\parallel}=0$ is zero. As the spins of the Ising phase are pinned out of the two-dimensional material plane by the Kane-Mele SOC, the in-plane spin-moment vanishes at leading order in $B_\parallel$, $M_{\parallel}^{\theta_v=0}=0$. Thus, the Ising-XY phase boundary evolves in $n_e-B_{\parallel}$ space according to
\begin{equation} \label{eq:dn_dB}
    \frac{\partial n_e}{\partial B_\parallel}=\frac{M_\parallel^{\theta_v=\pi/2}}{\mu^{\theta_v=\pi/2}-\mu^{\theta_v=0}}.
\end{equation}
Note $M_\parallel^{\theta_v=\pi/2}$ is the magnitude of the spin magnetization so the sign of $\frac{\partial n_e}{\partial B_\parallel}$ is determined by the denominator:
\begin{equation}
    \mu^{\theta_v=\pi/2}-\mu^{\theta_v=0}\equiv
\mu^{\theta_v=\pi/2}_b+\mu^{\theta_v=\pi/2}_{ex}-\mu^{\theta_v=0}_b-\mu^{\theta_v=0}_{ex}<0.
\end{equation}
We used Eq.~\ref{eq:mu_diff_band} and Eq.~\ref{eq:mu_diff_ex} to establish the above relationship. Fig.~\ref{fig:soc_6}.\textbf{b}) shows that the ratio $M_\parallel^{\theta_v=\pi/2}/(\mu^{\theta_v=\pi/2}-\mu^{\theta_v=0})$ has very weak dependence on the strength of SOC $\lambda$. Given that $M_\parallel^{\theta_v=\pi/2}=g\mu_B S_x^{\theta_v=\pi/2}$ and considering $S_x^{\theta_v=\pi/2}\approx 1$ per hole at the densities where $XY$ state are the groundstate, it is clear that the weak $\lambda$ dependence in $\frac{\partial n_e}{\partial B_\parallel}$ is attributed to the small increment of $|\mu^{\theta_v=\pi/2}-\mu^{\theta_v=0}|$ as $\lambda$ increases. 

Our analysis enables to compute the in-plane spin susceptibility $\chi_{\parallel}$ of the Ising-state $\ket{\theta_v=0,\theta_s=\pi}$:

\begin{equation}
   \chi_{_\parallel} = \frac{\partial^2 E^{\theta_v=0}}{ \partial B_{\parallel}^2} \sim 8~\mu  \text{eV~T}^{-2}.
\end{equation}
See Appendix B for more about in-plane spin susceptibility.
Since $1/\chi_{_\parallel}$ represents the resistance to collectively rotating the spin away from the uniaxial magnetic easy-axis, $1/\chi_{_\parallel}$ increases with the strength of spin-orbit coupling. 


%

\section{Quarter Metals in Bernal Bilayer Graphene and Beyond}

\begin{table}[t]
    \centering
    \begin{tabular}{|c|c|c|}
        \hline
         & \textbf{RTG}  & \textbf{BBG} \\
        \hline
         $n_e$& $-2.7\times 10^{11}$ cm$^{-2}$ & 
         $-1.0\times 10^{11}$ cm$^{-2}$
         \\
         \hline
         $E_{ex}/|n_e| $& $-4.645$ meV & $-2.349$ meV \\
        \hline
         $\langle \tau_x\rangle/|n_e|$& $1$ & $0.86$ \\
        \hline
        $\langle \tau_z\rangle/|n_e| $& $0$ & $0$ \\
        \hline
         $\tau_{z,\vec{k}}$ &  \includegraphics[width=3.3cm]{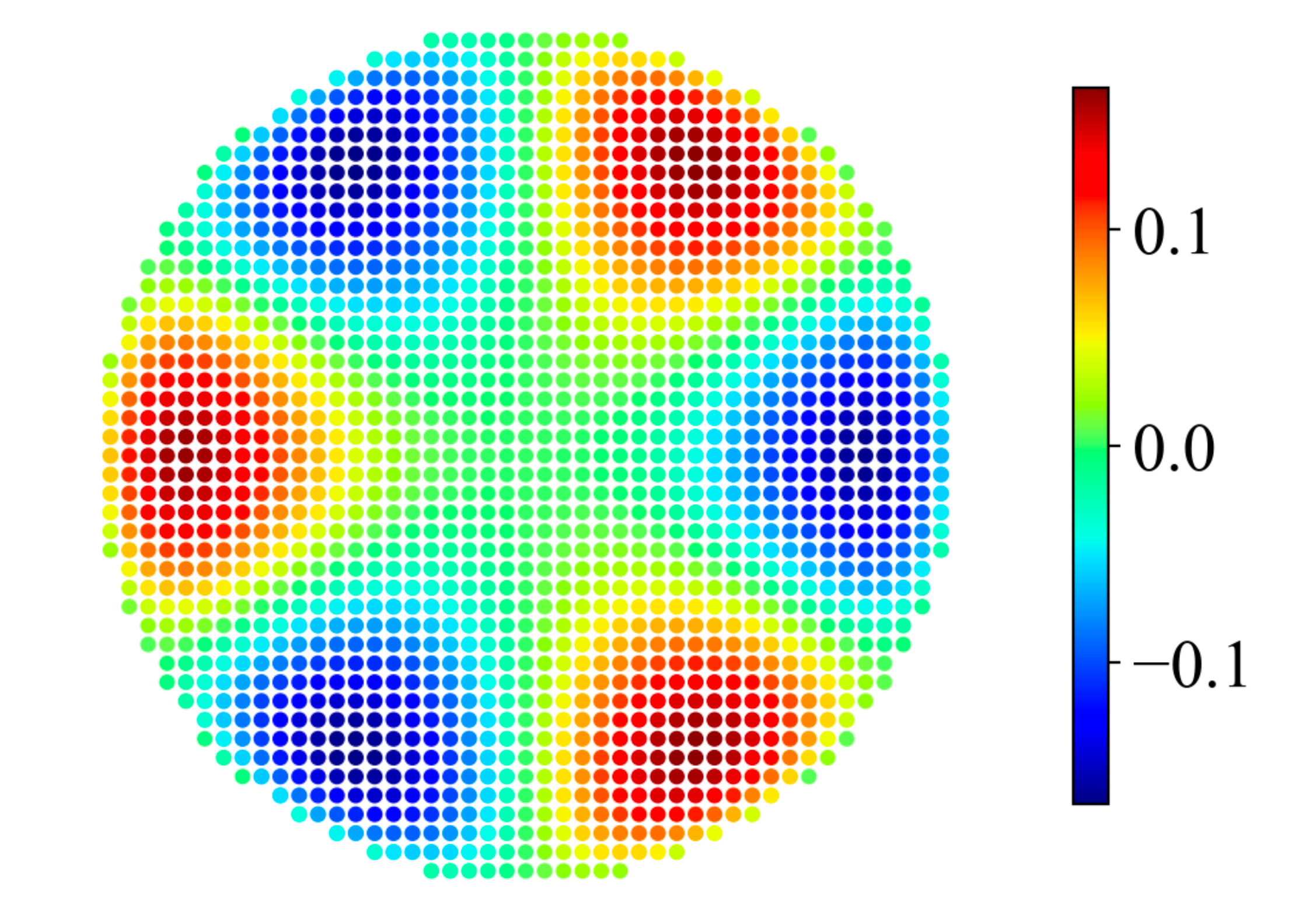} &  \includegraphics[width=3.3cm]{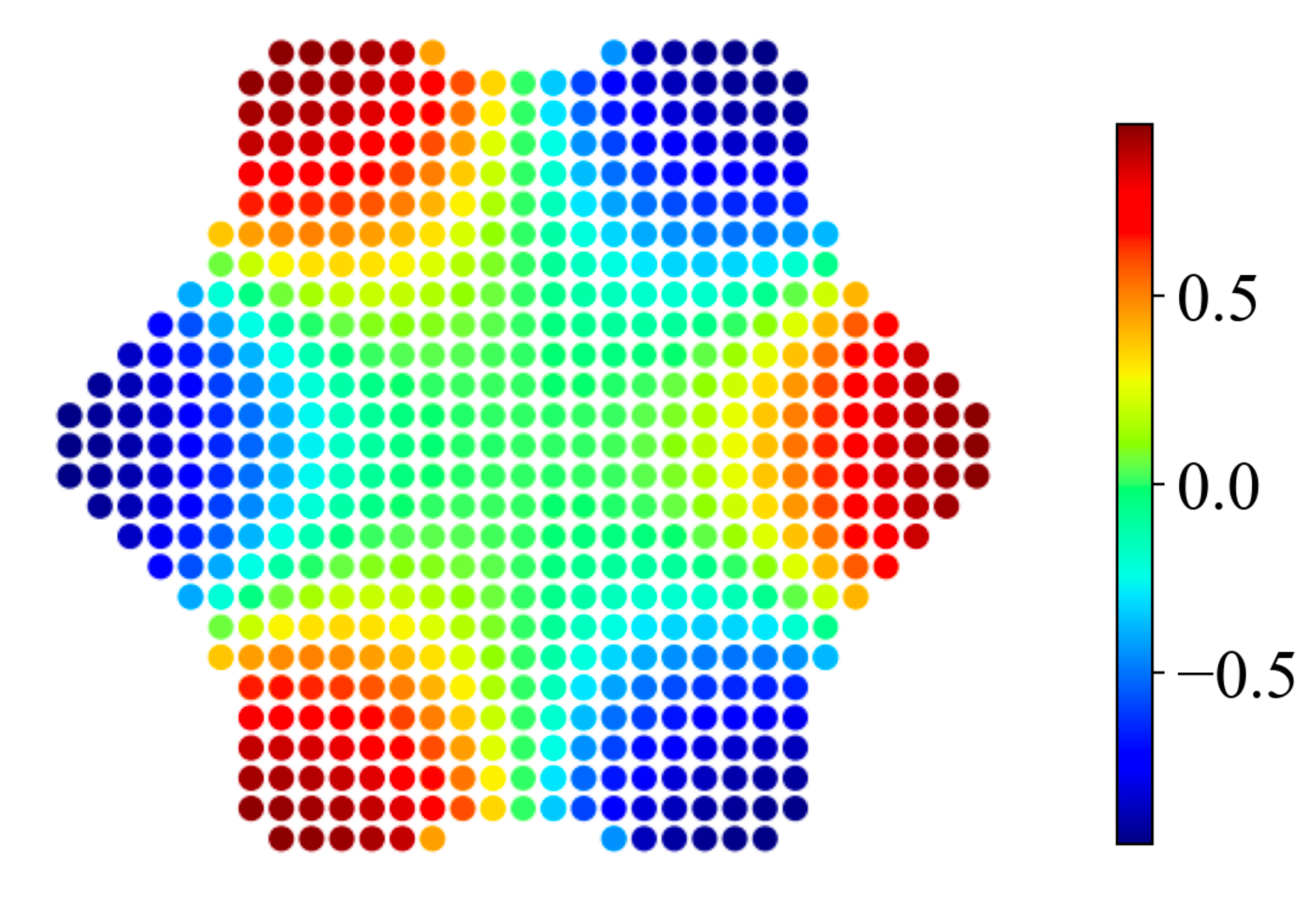} \\
        \hline
    \end{tabular}
    \caption{Comparison of the valley-XY quarter metal in rhombohedral trilayer graphene (RTG) and Bernal bilayer graphene (BBG). Here $n_e$, $E_{ex}/|n_e|$ and $\langle \tau_x \rangle$ represent density, the exchange energy per particle (measured respect to the paramagnetic state), and the order parameter, respectively. $\tau_{z,\vec{k}}=\langle \psi_{nk}|\tau_{z,k}|\psi_{nk}\rangle$ denotes the valley polarization of the hole-occupied Bloch states and $\langle\tau_z\rangle=\sum_{k}\tau_{z,\vec{k}}=0$. Note the stronger variation of $\tau_{z,\vec{k}}$ in BBG raises the exchange energy per hole in comparison to RTG. For these calculations, we set $U=28$ meV.}
    \label{tab:BBG}
\end{table}

So far, our study has primarily focused on the quarter metal phase diagram of rhombohedral trilayer graphene (RTG).
In this section, we contrasts the phase diagram of RTG with Bernal bilayer graphene (BBG). BBG is another system where experimental studies \cite{zhou_superconductivity_2021,de2022cascade,zhang2023enhanced} have confirmed the presence of the quarter metal. Our objective here is to elucidate the core principles of magnetism that could be universally applicable to all multilayer graphene electron gas systems, while also highlighting specific nuances that are harder to predict.

\begin{figure}[t]
    \centering
\includegraphics[width=\columnwidth]{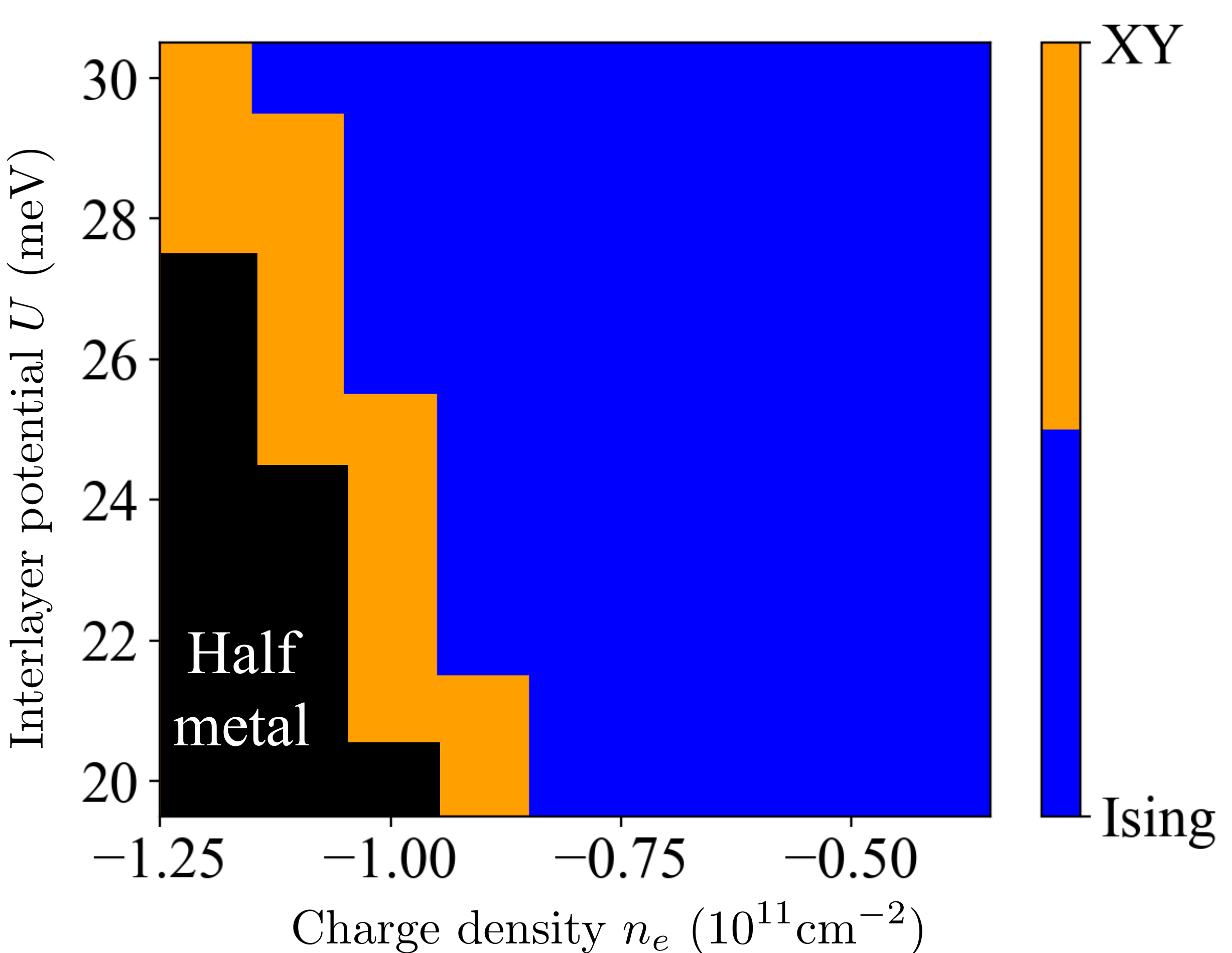}
    \caption{Mean-field phase diagram of quarter-metal phases in BBG shows valley-XY phase is the groundstate close to the half-metal. As the hole density decreases, the valley-XY state transitions into the valley-Ising state.}
    \label{fig:BBG}
\end{figure}

The mean-field phase diagram of BBG is shown in Fig.~\ref{fig:BBG}, which agrees qualitatively  with Ref.~\cite{PhysRevB.107.L201119}. Both RTG and BBG's quarter metal phase diagrams feature the valley-XY and valley-Ising states, yet they differ significantly in certain aspects. First, quarter metal in BBG occurs at density $|n_e|$ that is 3 to 5 times smaller than that in RTG. This is consistent with the experimental observations in Ref.~\cite{zhou_isospin_2021,zhou_half_2021}. Second, the valley-XY phase in BBG occupies a very small region in the $n_e-U$ phase diagram. Finally, there is a notable distinction in the magnetic anisotropic energy landscape: in RTG, the valley-Ising transitions into the valley-XY as $|n_e|$ decreases, whereas in BBG, the transition occurs from the valley-XY to the valley-Ising with decreasing $|n_e|$. We'll first address the initial two points using general principles inherent to multilayer graphene systems. Subsequently, we'll discuss the difference in magnetic anisotropic energy landscape, which exemplifies the nuanced specifics that challenge predictability.

Given that BBG's quarter metal is observed at a density roughly five times less than RTG, the area of the Fermi sea in BBG contracts by the same factor when compared to RTG. While this reduced Fermi sea area doesn't alter the spin and valley order parameters of the Ising phase, it does enhance the variation of valley-polarization within the Fermi sea for the valley-XY phase. This variation raises valley-XY's exchange energy rendering it less stable as the ground state. This underlines the reason why the valley-XY phase occupies a smaller $n_e-U$ phase space in BBG compared to RTG. 
To further elucidate this point, we plot the valley-polarization $\tau_z$ of the valley-XY phase in BBG in Table \ref{tab:BBG}, obtained from self-consistent mean-field calculation. Similar to RTG, as the Bloch states traverse the Fermi surface, the vector $\vec{\tau}_{\vec{k}}=(\tau_{x,\vec{k}}, \tau_{z,\vec{k}})$ exhibits oscillatory behavior reminiscent of a domain wall. Yet, a closer look at the color bar in Table \ref{tab:BBG} shows that the variations of $\tau_{z,\vec{k}}$ in BBG are more pronounced than in RTG.

We now turn our attention to the magnetic anisotropic energy landscapes of quarter metal in RTG and BBG. In both systems, the exchange energy invariably promotes the Ising phase as the ground state, while the band energy gives always preference to the XY phase. However, their mean-field energy differences are minimal and are swayed by detailed factors such as layer polarization.
For BBG, our observations indicate that, at large $|n_e|$, the band-energy difference is greater than the exchange-energy difference. This means the valley-XY phase is the dominant ground state. In contrast, RTG behaves oppositely.

\section{Summary}

In conclusion, we have addressed four pertinent questions concerning the low-density portion of the magnetic phase diagram in rhombohedral trilayer graphene:

\textit{1) Magnetic anisotropy of the quarter-metal groundstate:}
We identified two competing groundstate for the quarter metal: the valley-Ising state and the valley-XY state, detailed extensively in Sec.~II.  
At large values of $|n_e|$ and $U$, the valley-Ising state is the quarter metal groundstate. As we decrease  $|n_e|$ and $U$, it undergoes a first order phase transition to become the valley-XY state. This transition is driven by both the band and exchange energy. 
Although the exchange effect unambiguously favors the valley-Ising state as the ground state, the exchange energy of the valley-XY state becomes similar to the valley-Ising state with reducing $|n_e|$. This narrowing of exchange energy gap between them is due to the subtle differences in their layer polarization.



\textit{2) Sublattice pseudospin polarization and the re-entrance of paramagnetic state: } 
We found that the exchange energy difference between the quarter metal and the paramagnetic state, denoted as $E_{ex}^{\theta_v}-E_{ex}^{PM}$, can be strongly influenced by the layer-pseudospin. Given that the area of the Fermi sea in the quarter metal is 12 times larger than the Fermi pockets in the paramagnetic state, the Fermi level of the quarter metal is situated further from neutrality compared to that of the paramagnetic state.
As a result, as $|n_e|$ reduces, the layer polarization in the paramagnetic phase increases more rapidly than in the quarter metal phase, resulting in $\partial(E_{ex}^{\theta_v}-E_{ex}^{PM})/\partial n_e >0$.

\textit{3) Magnetic anisotropy and Kane-Mele spin-orbit coupling:}
We examined the influence of Kane-Mele spin-orbit coupling (SOC) on the quarter-metal phase diagram. 
This SOC term lowers the energy of a many-body state where the z-component of the spin-polarization ($s_z$) is anti-parallel to the z-component of the valley-polarization ($\tau_z$). Here $\hat{z}$ is the unit vector normal to the two-dimensional material plane.
In the case of valley Ising quarter metal, this SOC establishes a magnetic easy-axis. Here, the spin-quantization axis of all Bloch states are uniformly aligned along $\pm\hat{z}$ and that the valley-polarization is everywhere 
antiparallel to it, i.e~$\mp\hat{z}$.
For the valley-XY quarter metal, this SOC introduces a magnetic easy-plane. In this scenario, both the spin-polarization and valley-polarization of the Bloch states predominantly orient themselves perpendicular to $\hat{z}$. Minor oscillations in $\tau_z$ and $s_z$ are observed within the Fermi sea.
This anisotropy indicates that while the Kane-Mele SOC gives preference to the valley-Ising as the ground state, an in-plane magnetic field favors the valley-XY as the ground state.

\textit{4) Displacement of Ising-XY boundary:}
In our study of the Ising-XY boundary in the four dimensional parameter space --- defined by density $n_e$, interlayer potential $U$, in-plane magnetic field $B_{\parallel}$ and SOC strength $\lambda$ -- we found that the Ising-XY phase boundary follows the trajectory  $\partial n_e/\partial B_{\parallel} \sim -0.5\times 10^{11} \text{cm}^{-2}\text{T}^{-1}$ within the $n_e$ and $B_{\parallel}$ parameter space. Furthermore, we  have computed the in-plane spin susceptibility of the valley-Ising quarter-metal to be $\chi_{_\parallel}\sim 8~\mu\text{eV} ~\text{T}^{-2}$. 

We centered our attention on RTG due to its experimental significance and its representation as the most rudimentary form of rhombohedral-stacked multilayer graphene. It is noteworthy that the density-of-states (DOS) in RTG significantly exceeds that of its ABA Bernal-stacked trilayer graphene counterpart. Levering on this pronounced DOS, Ref.~\cite{han2023correlated,han2023orbital}  has recently identified correlated metals and insulators in rhombohedral-stacked pentalayer graphene even without the application of an external electric displacement field.


\textit{Note added:} During finalizing our work, we became aware of parallel experimental work focusing on determining the strength of Kane-Mele Spin-Orbit Coupling (SOC) using the effect of $B_\parallel$ on phase transition boundaries in RTG. In the quarter-metal region, two competing ground states have been identified: the valley-imbalanced (VI) orbital ferromagnet and the intervalley coherent (IVC) phase, which is equivalent to the valley Ising and valley XY phase, respectively, as discussed in this work. Ref. \cite{arp2023intervalley} provides $dn_e/dB_\parallel\sim -0.43\times 10^{11} \text{cm}^{-2}\text{T}^{-1}$, which is in rough agreement with our finding. This experiment predicts the Kane-Mele SOC strength $\sim 50~\mu$eV.

\textit{Acknowledgement:}
We acknowledgement insightful conversation with Nemin Wei, Andrea Young and Haoxin Zhou. We are grateful to the University of Kentucky Center for Computational Sciences and Information Technology Services Research Computing for their support and use of the Morgan Compute Cluster and associated research computing resources.

\newpage

\appendix

%
\section{Effect of inter-valley scattering on quarter metal groundstates}
The landscape of magnetic anisotropy energy can be affected by several independent parameters. In the main text, we have examined four such parameters: $n_e$, $U$, $\lambda$, and $B_\parallel$. In this section, we study the impact of inter-valley scattering on quarter-metal ground states and how it alters the $n_e$-$U$ phase diagram outlined in Fig.~\ref{fig:soc_2}. 
We introduce the the self-energy $\hat{\Sigma}_{\perp\vec{k}}$ into Eq.~\ref{eq:H_MF} of the maintext, where
\begin{align}   \hat{\Sigma}_{\perp\vec{k}}=u_\perp\sum_{a=x,y}\left[ \text{Tr}(\hat{\tau}_a \hat{\rho}_{\vec{k}}^{\theta_v})\hat{\tau}_a - \hat{\tau}_a \hat{\rho}_{\vec{k}}^{\theta_v} \hat{\tau}_a \right].
\end{align}
Here $u_\perp$ represents the strength of the inter-valley Coulomb scattering.
In the above equation, first term denotes the Hartree contribution, while the second one corresponds to the Fock contribution. Their contribution to the total energy is given by the following:
\begin{align}
    E_{\perp H}^{\theta_v} &=\frac{u_\perp}{2} \sum_{a=x,y} \int \frac{d^2k}{(2\pi)^2} \left[ \text{Tr}(\hat{\tau}_a\hat{\rho}_{\vec{k}}^{\theta_v})\right]^2 ,\\
    E_{\perp ex}^{\theta_v} &=-\frac{u_\perp}{2} \sum_{a=x,y}\int \frac{d^2k}{(2\pi)^2} \text{Tr}\left[ \hat{\tau}_a \hat{\rho}_{\vec{k}}^{\theta_v}\hat{\tau}_a \hat{\rho}_{\vec{k}}^{\theta_v}\right],
\end{align}
where $\hat{\rho}_{\vec{k}}^{\theta_v}$
is the density-matrix determined self-consistently. As shown in Table \ref{tab:ivs_components},
despite a notable exchange energy shift related to the Dirac sea,  the energy difference between them is very small:$
E_{\perp,H}^{\theta_v=0} 
+E_{\perp,ex}^{\theta_v=0} 
-E_{\perp,H}^{\theta_v=\pi/2} 
-E_{\perp,ex}^{\theta_v=\pi/2} \sim 4\mu$eV per-hole. Furthermore, Table \ref{tab:energy_diff} shows that the variations in both band energy and (long-range Coulomb )exchange energy, due to $u_\perp$, are similar for both the Ising and XY states.
Consequently, the introduction of inter-valley scattering doesn't bring about significant alterations to the quarter-metal phase diagram.

\begin{table}
    \centering
    \begin{tabular}{|c|c|c|}
        \hline
        $\mathbf{\theta_v}$ & $\mathbf{0}$ & $\mathbf{\pi/2}$ \\
        \hline
        $\mathbf{\Delta E_b /|n_e|}$ (meV) & $0.761$ & $0.753$ \\
        \hline
        $\mathbf{\Delta E_{ex}/|n_e|}$ (meV) & $-0.713$ & $-0.703$ \\
        \hline
    \end{tabular}
    \caption{Change in band and Fock exchange energy for valley-Ising and valley-XY phase due to change in $u_\perp$ from $0$ meV to $0.5$ meV at $(n_e,U)=-2.9\times 10^{11}~\text{cm}^{-2},18~\text{meV}$.}
    \label{tab:energy_diff}
\end{table}

\begin{table}
    \centering
    \begin{tabular}{|c|c|c|}
        \hline
        $\mathbf{\theta_v}$ & $\mathbf{0}$ & $\mathbf{\pi/2}$ \\
        \hline
        $\mathbf{E_{\perp H}/|n_e|}$ (meV) & $0$ & $0.41$ \\
        \hline
        $\mathbf{E_{\perp ex}/|n_e|}$ (meV) & $-55.814$ & $-56.228$ \\
        \hline
    \end{tabular}
    \caption{Hartree and exchange energy component of intervalley scattering for valley-Ising and valley-XY phase using $u_\perp=0.5$ meV at $(n_e,U)=-2.9\times 10^{11}~\text{cm}^{-2},18~\text{meV}$.}
    \label{tab:ivs_components}
\end{table}


\section{Spin susceptibility of valley-Ising quarter metal}


\begin{figure}[ht]
    \centering
    \includegraphics[width=0.9\columnwidth]{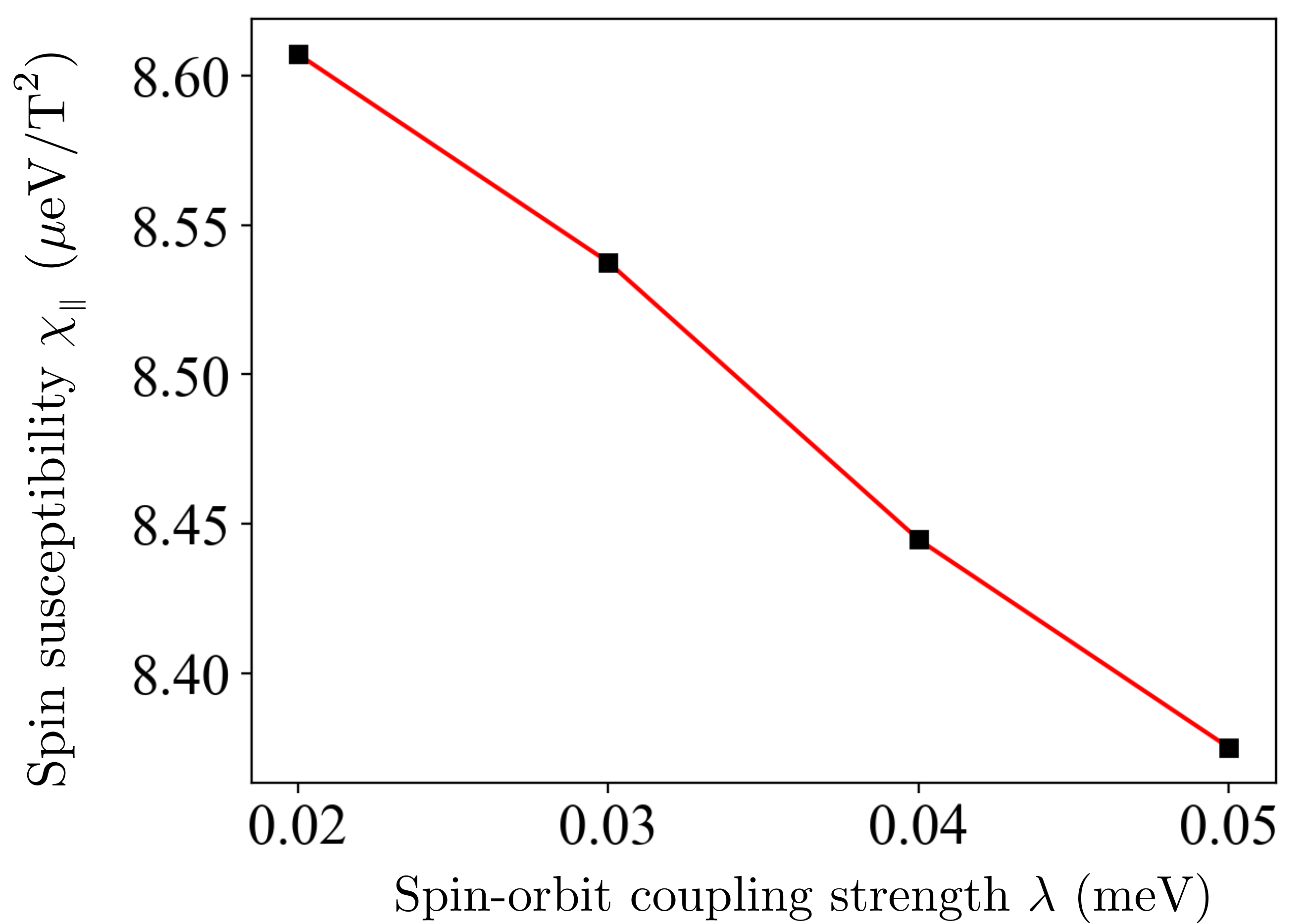}
    \caption{ In-plane spin susceptibility of valley-Ising quarter metal is plotted against increasing SOC strength $\lambda$ for $n_e,U=-2.9\times 10^{11}~\text{cm}^{-2},18~\text{meV}$.}
    \label{fig:soc_9}
\end{figure}

We compute the in-plane spin susceptibility ($\chi_{_\parallel}$) of the valley-Ising quarter metal for different values of spin-orbit coupling ($\lambda$). $\chi_{_\parallel}$ is defined as:
\begin{align}
\chi_{_\parallel}=\frac{\partial^2 E^{\theta_v=0}}{\partial B_\parallel^2}\left. \vphantom{\frac{A}{B}} \right|_{B_\parallel=0}=\frac{\partial M_{\parallel}^{\theta_v=0}}{\partial B_\parallel}\left. \vphantom{\frac{A}{B}} \right|_{B_\parallel=0}
\end{align}
 Fig.~\ref{fig:soc_9} shows $\chi_{_\parallel}$ $v.s.$ $\lambda$ at $n_e,U=-2.9\times 10^{11}~\text{cm}^{-2},18~\text{meV}$. 
 Since the SOC pinned the spin-quantization axis of the valley-Ising state in the direction normal to the applied in-plane field $B_{\parallel}$, 
$\chi_{_\parallel}$ decreases with increasing $\lambda$.

\section{Energy landscape of quarter-metal in the presence of spin-orbit coupling}
\begin{figure*}[t]
    \centering
    \includegraphics[width=2\columnwidth]{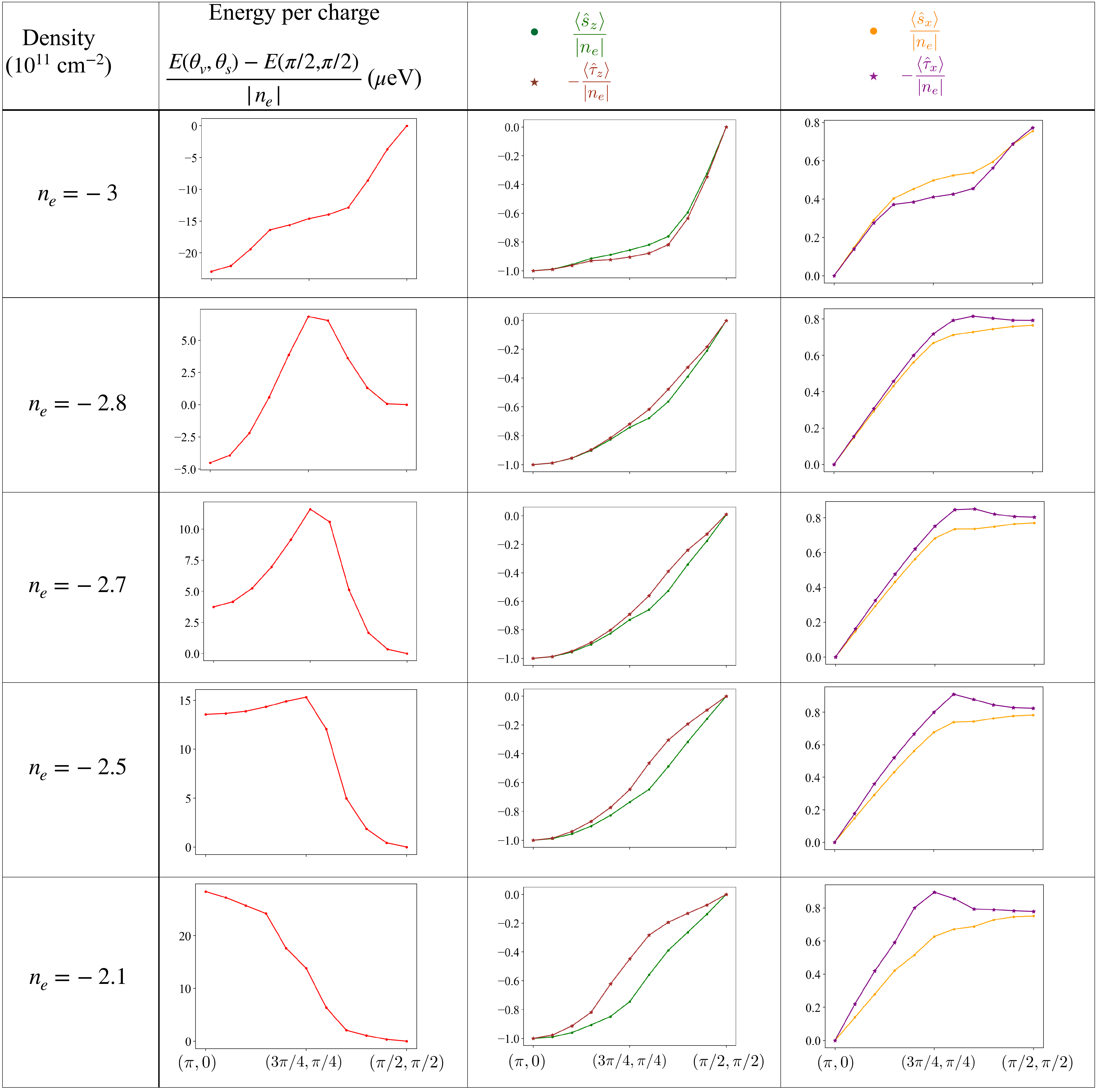}
    \caption{The first column shows the total energy v.s.~$\theta$ while the second and third columns show the order parameters of the converged solution $v.s.$~$\theta$. $\theta$ is defined in Eq.~C1 and and we set the $\lambda_L=1~\mu$eV, $\lambda=30~\mu$eV.
    }
    \label{fig:soc_10}
\end{figure*}
 So far, we have identified only two quarter metal groundstate: valley-Ising and valley-XY phases. They are separated by a first order phase transition boundaries in the $n_e-U$ space. In this section, we used the following Lagrange multipliers to further examine their energy landscape:
\begin{align}
    \hat{H}_L(\vec{n}) =& \lambda_L[\vec{n}\cdot \hat{\vec{s}}+(\vec{n}\times \vec{z})\cdot \hat{\vec{\tau}}]\notag\\ =& \lambda_L (\text{sin}\theta~ \hat{s}_x+\text{cos}\theta~ \hat{s}_z+\text{sin}\theta~ \hat{\tau}_x -\text{cos}\theta~ \hat{\tau}_z).
\end{align}
Here $\vec{n}=(\sin\theta,\cos\theta)$ represents the angle of spin-polarization. We opted not to delve into the two-dimensional energy landscape generated by independent valley and spin rotations as $H_{KM}=\lambda \sigma_zs_z\tau_z$ favors the anti-alignment of the $s_z$ with $\tau_z$.  Note $\lambda\langle\sigma_z\rangle>0$.

 In Fig.~\ref{fig:soc_10}, we plot the evolution of both the energy and the associated spin and valley order parameters for the quarter-metal phases as functions of hole density. At a high density of $n_e=-3\times 10^{11}~\text{cm}^{-2}$, the valley-Ising state, denoted by $\ket{\theta_v=\pi,\theta_s=0}$, establishes itself as the global minimum, whereas the valley-XY state $\ket{\theta_v=\pi/2,\theta_s=\pi/2}$ appears unstable. However, when the density is lowered to $n_e=-2.8\times 10^{11}~\text{cm}^{-2}$, the valley-XY state stabilizes and becomes a local minimum. Reducing the density further to $n_e=-2.7\times 10^{11}~\text{cm}^{-2}$ prompts a shift in the global minimum from $\ket{\theta_v=\pi,\theta_s=0}$ to $\ket{\theta_v=\pi/2,\theta_s=\pi/2}$, signifying a first-order phase transition. At  $n_e=-2.1\times 10^{11}~\text{cm}^{-2}$,  $\ket{\theta_v=\pi,\theta_s=0}$ becomes unstable. Note that the Lagrange multiplier $\lambda_L$ has a minimal impact on shifting the Ising-XY phase boundary.

\bibliography{references}

\end{document}